\documentclass{ws-ijmpcs}

\graphicspath{{figures/}{fig/}}

\newcommand{\ben}{\begin{eqnarray}}
\newcommand{\een}{\end{eqnarray}}
\newcommand{\nnu}{\nonumber\\}
\newcommand{\bef}{\begin{figure}[!htp]}
\newcommand{\eef}{\end{figure}}

\newcommand{\bea}{\begin{eqnarray}}
\newcommand{\eea}{\end{eqnarray}}

\def\ba{\begin{linenomath*}\begin{equation}}
\def\ea{\end{equation}\end{linenomath*}}

\def\trimmarks{}

\begin{document}

\markboth{Y.Q. Ma \& J.W. Qiu}
{QCD Factorization and PDFs from Lattice QCD}

%
\catchline{}{}{}{}{}
%

\title{QCD Factorization and PDFs from Lattice QCD Calculation\footnote{Presented by J.-W. Qiu.}
}

\author{Yan-Qing Ma$^a$ and Jian-Wei Qiu$^{b}$}

\address{$^a$Maryland Center for Fundamental Physics,
University of Maryland, \\
College Park, Maryland 20742, USA
\\
Center for High-Energy Physics, Peking University, Beijing, 100871, China}

\address{$^b$Physics Department,
		Brookhaven National Laboratory,
		Upton, NY 11973-5000, U.S.A.
\\
		C.N.\ Yang Institute for Theoretical Physics
		and Department of Physics and Astronomy,\\
             	Stony Brook University,
             	Stony Brook, NY 11794-3840, USA
\\
yqma@bnl.gov, jqiu@bnl.gov}

\maketitle

\begin{history}
\received{November 27, 2014 }
\revised{Day Month Year}
\published{Day Month Year}
\end{history}

\begin{abstract}
In this talk, we review a QCD factorization based approach to extract
parton distribution and correlation functions from lattice QCD calculation
of single hadron matrix elements of quark-gluon operators.
We argue that although the lattice QCD calculations
are done in the Euclidean space, the nonperturbative collinear behavior
of the matrix elements are the same as that in the Minkowski space,
and could be systematically factorized into parton distribution functions
with infrared safe matching coefficients.  The matching coefficients
can be calculated perturbatively by applying the factorization formalism
on to asymptotic partonic states.
\end{abstract}

\ccode{PACS numbers:\ 12.38.Bx, 13.88.+e, 12.39.-x, 12.39.St}

\section{Introduction}	
\label{sec:intro}

Parton distribution and correlation functions describe the fascinating relation
between a hadron and the quarks and gluons (or partons) within it.
They carry rich information on hadron's mysterious partonic structure that
cannot be calculated by QCD perturbation theory.
Parton distribution functions (PDFs) are the simplest of all correlation functions,
$f_{i/h}(x,\mu^2)$, defined as the probability distributions to find a quark,
an antiquark, or a gluon ($i=q,\bar{q},g$) in a fast moving hadron
to carry the hadron's momentum fraction between $x$ and $x+dx$,
probed at the factorization scale $\mu$.
They carry an invaluable source of information on the partonic structure
and the confinement-sensitive partonic dynamics of colliding hadron(s), and they also
play an essential role to connect the cross sections of colliding hadron(s) to 
the short-distance scattering between quarks and gluons.
Without them, we would not be able to understand the hard probes,
cross sections with large momentum transfers, in high energy hadronic collisions,
as well as the discovery of Higgs particles in proton-proton collisions at the LHC.
PDFs are nonperturbative, but, universal, and have been traditionally
extracted from QCD global analysis of all existing high energy scattering data
in the framework of QCD factorization\cite{Gao:2013xoa}\cdash\cite{Alekhin:2013nda}.

Unlike cross sections, PDFs are not direct physical observables.
Their extraction from data on hard probes in high energy scattering always
rely on the validity of QCD factorization\cite{Collins:1989gx} or the approximation
to neglect all contributions suppressed by inverse powers of the large momentum transfer.
On the other hand, PDFs are hadronic matrix elements of well-defined operators,
which are made of quark or gluon fields in QCD, along with a proper ultraviolet
renormalization specified for these composite operators\cite{Collins:1981uw}.
It might be possible to derive these PDFs directly from {\it ab initio} calculations of QCD,
such as lattice QCD, and to compare the calculated PDFs with the extracted ones. 
However, it is extremely difficult to calculate PDFs
in lattice QCD since the PDFs are defined by hadronic matrix elements
of non-local operators in the Minkowski space, while all lattice QCD
calculations are done with an Euclidean time.
The moments of PDFs, $\int_0^1 dx x^{n-1} f_{i/h}(x,\mu^2)$,
given by the hadronic matrix elements of local operators, 
have been traditionally investigated by the lattice QCD community.
Although progress has been made, the comparison between the lattice calculations
and the moments of extracted PDFs has not reached to a satisfactory
accuracy\cite{Dolgov:2002zm,Gockeler:2004wp}.  

Recently, Ji\cite{Ji:2013dva} introduced a set of quasi-PDFs,
defined in terms of hadronic matrix elements of equal time correlators,
calculable in lattice QCD\cite{Lin:2014zya},
and suggested that the quasi-PDFs become the normal PDFs
when the hadron momentum $P_z$ is boosted to the infinity.
However, since the hadron momentum in lattice QCD calculation
is effectively bounded by the lattice spacing,
the $P_z\to\infty$ limit is hard to achieve in lattice calculations, Ji introduced the
large-momentum effective field theory of QCD\cite{Ji:2014gla} and suggested 
that it provides a frame work to evaluate the difference 
between PDFs and quasi-PDFs due to a finite $P_z$.
The connection between the PDFs and quasi-PDFs
is further complicated by the fact that the operator defining the
quasi-PDFs are power ultra-violet (UV) divergent,
while the operators defining the normal
PDFs have only logarithmic UV divergence.

In this talk, we review a QCD factorization approach, 
proposed recently by us\cite{Ma:2014jla}, for extracting PDFs 
from lattice QCD calculations of single hadron matrix elements of quark-gluon correlators.
For all QCD factorization treatments of hadronic cross sections
with large momentum transfer(s), PDFs were introduced to absorb all leading power 
partonic collinear (CO) divergences associated with the colliding hadron(s).
For example, all leading power partonic CO divergences of the lepton-hadron 
deep inelastic scattering (DIS) cross section are universal and can be completely 
absorbed into the PDFs, if we neglect the corrections suppressed by 
the inverse powers of the large momentum transfer $Q=\sqrt{-q^2}$, 
the virtuality of the exchange photon of momentum $q$.  
All leading power partonic CO divergences of the DIS cross sections,
which is proportional to a single hadron matrix element of two conserved 
electromagnetic currents, $\langle h(P)| j_\mu(\xi) j_\nu(0)|h(P)\rangle$, 
come from the region of phase space where all active partons' transverse 
momenta, $k_{i\perp}^2\to 0$, where $i=1,2,\dots$, with respect to 
the colliding hadron's momentum $P^\mu=(P_0,0_\perp, P_z)$. 
Our proposal is based on the observation that the leading power CO divergences 
in the limit $k_{i\perp}^2\to 0$ are the same regardless if the time defining the 
hadronic matrix element is in the Minkowski or the Euclidean space.  
Our method involves four steps: 
1) identify hadronic matrix elements that are both calculable in lattice QCD 
and factorizable into PDFs -- referred as lattice ``cross sections'', 
2) generate the ``data'' of the lattice ``cross sections'', 
3) evaluate the factorized coefficient functions between the lattice ``cross sections''
and the PDFs, and 
4) perform the global analysis of the lattice ``data'' using the factorization formalisms 
to extract the PDFs.  

Our approach requires a large virtuality -- the probe's momentum transfer, 
$Q\gg \Lambda_{\rm QCD}$, to ensure the validity of the QCD factorization 
of the lattice ``cross sections'' into PDFs.  The QCD factorization in our approach 
is an approximation to neglect corrections suppressed by the inverse powers of $Q$, 
which is not the same as the expansion in $1/P_z$ in Ji's approach.  
In our proposed approach, the hadron momentum $P_z$ is a large, but finite,
``observed'' momentum scale for the lattice ``cross sections'', 
similar to the collision energy $\sqrt{S}$ for hadronic cross sections, 
and is of the order of $Q$ or larger.  Our approach is effectively the same as 
the leading power QCD factorization approach for extracting PDFs from 
data of hadronic cross sections, except the cross sections are replaced 
by the lattice ``cross sections'' evaluated in the Euclidean space.

With the limitation of current lattice size and computing power, it is certainly
too expensive, or even impossible to calculate PDFs at small $x$.  However,
lattice calculations could certainly provide very valuable information on PDFs 
in the valence region, in particular, in the regime when $x\to 1$, 
where the accuracy of experimental data is limited\cite{Peng:2014hta}, 
while theoretical techniques to resum large $x$ perturbative contribution 
have been developed and improved.  
Furthermore, with lattice QCD calculations of the
hadronic matrix elements of the same operators on meson states, or states
of baryons other than the proton, this QCD factorization approach could  
in principle extract PDFs of mesons or exotic baryons and their partonic structure, 
without performing high energy scattering on mesons and exotic baryons, 
which could be very difficult if not impossible. 

\section{Lattice ``cross sections'' and factorization}
\label{sec:factorization}

We define a lattice ``cross section'', 
$\widetilde{\sigma}_{h,\text{E}}^\text{Lat}(\tilde{x},1/a,P_z)$, 
as the Fourier transform of a single hadron matrix element, 
$\langle h(P)|{\cal O}(\psi,A)|h(P)\rangle$, with the colliding hadron momentum $\vec{P}$ 
along $z$-direction and large, $P^0 \approx |P_z| \gg \Lambda_{\rm QCD}$, and 
an operator ${\cal O}(\psi,A)$ of quark $\psi$ and/or gluon $A$ field, 
where the transverse lattice spacing $a$ defines the hard scale $\sim 1/a$, 
and the dimensionless parameter $\tilde{x}$, defined below, and 
$P_z$ mimics the ``rapidity'' and ``collision energy'' of the ``cross section'', respectively.  
A good lattice ``cross section'', $\widetilde{\sigma}_{h,\text{E}}^\text{Lat}(\tilde{x},1/a,P_z)$, 
should have the following properties\cite{Ma:2014jla}:
\begin{itemize}
\item
It must be calculable in lattice QCD with an Euclidean time, 
indicated by the superscript ``Lat'' and the subscript ``E'',
\item
It is infrared (IR) safe if it is calculated in lattice perturbation theory,
\item
All CO divergences of its continuum limit ($a\to 0$) can be factorized into the PDFs 
with perturbatively calculable hard coefficient functions.
\end{itemize}
Lattice QCD is a UV finite theory, and the lattice ``cross section'' calculated 
in its perturbation theory with a finite $a$ is also UV finite.  
However, the perturbatively calculated lattice ``cross section'', which is needed for 
extracting the finite coefficient functions when it is factorized into PDFs, 
might be UV sensitive -- perturbatively unstable when the lattice spacing $a$ 
is sufficiently small, if the operator ${\cal O}(\psi,A)$ defining the lattice ``cross section'' 
does not have a renormalizable continuum limit.  That is, a good lattice ``cross section''
also requires the operator to define its single hadron matrix element to have
a renormalizable continuum limit.

Once we identify good lattice ``cross sections'', we are able to factorize them in terms of the PDFs,
\begin{eqnarray}
\widetilde{\sigma}_{h,\text{E}}^\text{Lat}(\tilde{x},\frac{1}{a},P_z)
\approx
\sum_i \int_0^1 \frac{dx}{x}\,f_{i/h}(x,\mu^2)\,\widetilde{\cal C}_i(\frac{\tilde{x}}{x}, \frac{1}{a},\mu^2,P_z),
\label{eq:fac-direct}
\end{eqnarray}
where $\widetilde{\cal C}$'s are perturbative coefficient functions.  
By applying Eq.~(\ref{eq:fac-direct}) to various parton states, 
$|h(P)\rangle \to |f(P)\rangle$ with flavor $f=q,\bar{q},g$, 
the $\widetilde{\cal C}$'s can be systematically derived by calculating 
$\widetilde{\sigma}_{f,\text{E}}^\text{Lat}(\tilde{x},1/a,P_z)$ on a parton state 
$f$ in lattice QCD perturbation theory and $f_{i/f}(x,\mu^2)$ 
of the same parton state in perturbative QCD.   
As explained in the last section, we can systematically extract the PDFs from ``data'' of lattice 
``cross sections'' by using the factorization relation in Eq.~(\ref{eq:fac-direct}), and
perturbatively calculated coefficient functions.  The accuracy of the extracted PDFs could be 
improved perturbatively by more accurate coefficient functions, $\widetilde{\cal C}$'s.

Our strategy to search for good lattice ``cross sections'' could be summarized 
by the following schematic plot,
\begin{eqnarray}
\begin{split}
\widetilde{\sigma}_{h,\text{E}}^\text{Lat}(\tilde{x},1/a,P_z)
~\overset{\cal Z}\longleftrightarrow~
\tilde{\sigma}_{h,\text{E}}(\tilde{x},&\tilde{\mu}^2,P_z)
\label{eq:matching}\\
&\Updownarrow\\
\tilde{\sigma}_{h,\text{M}}(\tilde{x},&\tilde{\mu}^2,P_z)
~\overset{\cal C}\longleftrightarrow
~f_{i/h}(x,\mu^2)\, ,
\end{split}
\label{eq:matchings}
\end{eqnarray}
where $\tilde{\sigma}_{h,\text{E}}(\tilde{x},\tilde{\mu}^2,P_z)$ is 
the Euclidean space continuum limit of 
$\widetilde{\sigma}_{h,\text{E}}^\text{Lat}(\tilde{x},1/a,P_z)$ 
with a proper UV counter term (UVCT) to renormalize its UV divergence, 
if there is any, at a hard scale $\tilde{\mu}$; and 
$\tilde{\sigma}_{h,\text{M}}(\tilde{x},\tilde{\mu}^2,P_z)$ 
is the Minkowski space version of $\tilde{\sigma}_{h,\text{E}}(\tilde{x},\tilde{\mu}^2,P_z)$, 
as indicated by its subscript ``M''.   
If the operator ${\cal O}(\psi,A)$ is time-independent, 
we expect that $\tilde{\sigma}_{h,\text{E}}(\tilde{x},\tilde{\mu}^2,P_z)
= \tilde{\sigma}_{h,\text{M}}(\tilde{x},\tilde{\mu}^2,P_z)$.  
To show the factorization in Eq.~(\ref{eq:fac-direct}) is effectively to 
prove the factorization between $\tilde{\sigma}_{h,\text{M}}(\tilde{x},\tilde{\mu}^2,P_z)$
and $f_{i/h}(x,\mu^2)$ in the continuous Minkowski space, and to verify 
the matching between $\widetilde{\sigma}_{h,\text{E}}^\text{Lat}(\tilde{x},1/a,P_z)$ 
and $\tilde{\sigma}_{h,\text{E}}(\tilde{x},\tilde{\mu}^2,P_z)$.
If the composite operator ${\cal O}(\psi,A)$ is made of conserved currents/tensors, 
we do not need the UVCT($\tilde{\mu}^2$)
to define the $\tilde{\sigma}_{h,\text{E}}(\tilde{x},\tilde{\mu}^2,P_z)$, and we only need 
the bottom matching relation with $\tilde{\mu}=1/a$.  
It is therefore very important to show the bottom factorization relation in Eq.~(\ref{eq:matchings})
for any potential lattice ``cross sections'' in the Minkowski space,
\begin{eqnarray}
\tilde{\sigma}_{h,\text{M}}(\tilde{x},\tilde{\mu}^2,P_z)
\approx
\sum_i \int_0^1 \frac{dx}{x}\,f_{i/h}(x,\mu^2)\,{\cal C}_i(\frac{\tilde{x}}{x}, \tilde{\mu}^2,\mu^2,P_z)
\label{eq:factorization}
\end{eqnarray}
with corrections suppressed by the inverse powers of factorization scale $\mu$.

This factorization approach for extracting PDFs could be generalized for extracting 
other parton distribution and correlation functions.  
For example, for extracting transverse momentum dependent PDFs (TMDs), the lattice
``cross sections'' are necessarily to have dependence on additional momentum scale(s) 
different from the hard scale $\tilde{\mu}$\cite{MaQiu2014}.

\section{Case study: the quasi-PDFs }
\label{sec:quasiPDF}

As a case study, we discuss if the quasi-PDFs, introduced by Ji\cite{Ji:2013dva}, 
could be good lattice ``cross sections'' for extracting the PDFs.  

\subsection{Definition}
\label{sec:def}

The quasi-quark distribution of a hadron $h$ of momentum $P^\mu$ is defined as\cite{Ji:2013dva}
\begin{eqnarray}
\label{eq:ftq}
\tilde{f}_{q/h}(\tilde{x},\tilde{\mu}^2,P_z)
\equiv
\int \frac{d\xi_z}{2\pi} e^{-i \tilde{x}P_z \xi_z} \tilde{F}_{q/h}(\xi_z,\tilde{\mu}^2,P_z),
\end{eqnarray}
where $\tilde{F}_{q/h}(\xi_z,\tilde{\mu}^2,P_z)
=\langle h(P)| \overline{\psi}(\xi_z)\,\frac{\gamma_z}{2}\Phi_{n_z}^{(f)}(\{\xi_z,0\})\, 
\psi(0) | h(P) \rangle+\text{UVCT}(\tilde{\mu}^2)$ and $\xi_0=\xi_\perp=0$. 
Similarly, the quasi-gluon distribution is defined as
\begin{eqnarray}
\label{eq:ftg}
\tilde{f}_{g/h}(\tilde{x},\tilde{\mu}^2,P_z)
\equiv
\frac{1}{\tilde{x}P_z}
\int \frac{d\xi_z}{2\pi} e^{-i \tilde{x} P_z \xi_z} \tilde{F}_{g/h}(\xi_z,\tilde{\mu}^2,P_z),
\end{eqnarray}
where $\tilde{F}_{g/h}(\xi_z,\tilde{\mu}^2,P_z)=\langle h(P)| F_{z}^{\ \nu}(\xi_z)\,
\Phi_{n_z}^{(a)}(\{\xi_z,0\})\, F_{z\nu}(0) | h(P) \rangle +\text{UVCT}(\tilde{\mu}^2)$ 
with $\nu$ summing over transverse directions.  
In Eqs.~(\ref{eq:ftq}) and (\ref{eq:ftg}), $\tilde{\mu}$ is a renormalization scale, 
and the gauge links $\Phi_{n_z}^{(f,a)}(\{\xi_z,0\}) = 
{\cal P}\text{exp}[-ig\int_0^{\xi_z} d\eta_z\, A^{(f,a)}_z(\eta_z)]$ 
where ${\cal P}$ indicates the path ordering, the superscripts, ``$f$'' and ``$a$'', 
represent the fundamental and adjoint representation of SU(3) color of QCD, respectively, 
and $n_z^\mu = (0,0_\perp, 1)$, $n_z^2 = -1$ and $v\cdot n_z = -v_z$ for any vector $v^\mu$.
Since the operators defining these single hadron matrix elements have no explicit time dependence, 
the quasi-PDFs could be calculated in lattice QCD\cite{Ji:2013dva}. 

To show that these quasi-PDFs could be factorized into the PDFs 
as in Eq.~(\ref{eq:factorization}), we need to demonstrate that they are IR safe, 
UV renormalizable with the UVCTs, and all their perturbative CO divergences 
can be absorbed into the PDFs.  Following effectively the same arguments 
used in Ref.~[\refcite{Collins:1981uw}], it is straightforward to show that 
these quasi-PDFs of an asymptotic parton state are indeed IR safe. 
\begin{figure}[h]
\begin{center}
\begin{minipage}[c]{0.8in}
\includegraphics[width=0.8in]{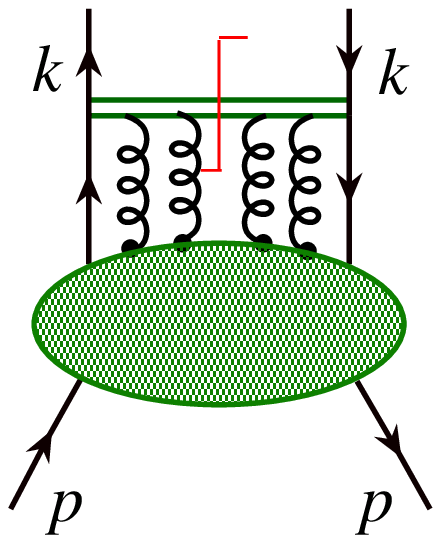}
\end{minipage}
= \
\begin{minipage}[c]{0.6in}
\includegraphics[width=0.6in]{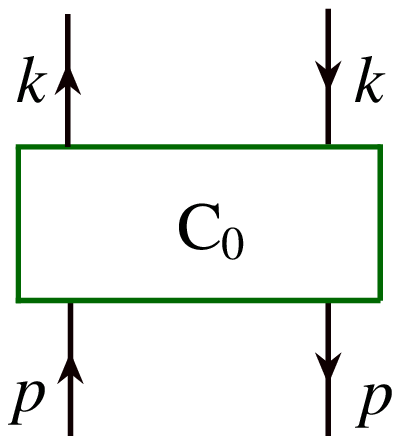}
\end{minipage}
\ + \
\begin{minipage}[c]{0.6in}
\includegraphics[width=0.6in]{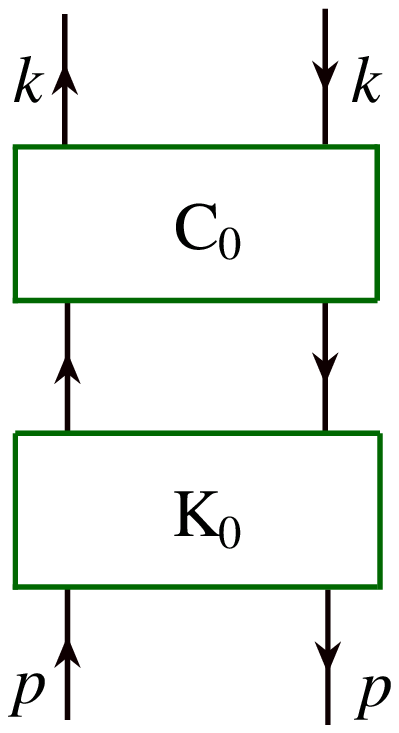}
\end{minipage}
\ + \
\begin{minipage}[c]{0.6in}
\includegraphics[width=0.6in]{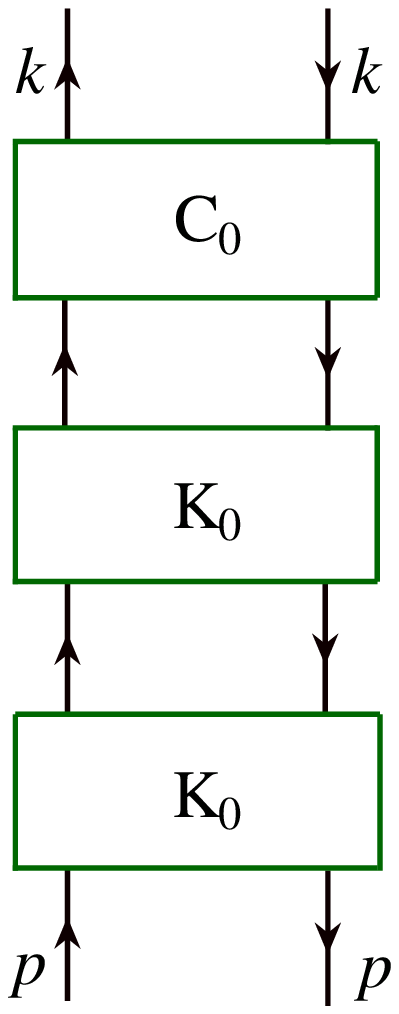}
\end{minipage}
\ + ...
\caption{\label{fig:quasi-quark}
Ladder expansion of the quasi-quark distribution.}
\end{center}
\end{figure}

\subsection{Factorization of CO divergences}
\label{sec:fac-co}

Factorization of the partonic CO divergences could be best demonstrated 
in the light-cone gauge, $n\cdot A=0$, with $n^\mu=(n^+,n^-,n_\perp)=(0,1,0_\perp)$.
In this gauge, for example, the leading power flavor non-singlet contribution to 
the quasi-quark distribution of a parton state of momentum $p$ 
can be approximated by a sum of ladder diagrams, as shown in Fig.~\ref{fig:quasi-quark}, 
plus UVCTs\cite{Ma:2014jla,Ellis:1978ty}, 
where $C_0$ and $K_0$ are two-particle irreducible (2PI) kernels.  
By definition, $K_0$ includes the two quark propagators connecting to the kernel above. 
We can introduce corresponding renormalized 2PI kernel $K$, with local UV divergences 
removed by the counter-terms of renormalized QCD Lagrangian.
Since the renormalized 2PI kernels with fixed external momenta are finite\cite{Ellis:1978ty},
all CO divergences of the ladder diagrams in Fig.~\ref{fig:quasi-quark} 
come from the integration of the loop momentum $k_i$ between two neighboring 2PI kernels,
and corresponding CO divergences are logarithmic. 
To factorize all leading power CO divergences of the quasi-quark distribution of a parton state 
into the PDFs of the same parton state, we introduce a projection operator, 
$\widehat{\cal P}$ to act on the kernel $K$ so that $\widehat{\cal P}K$ 
picks up the leading logarithmic CO divergence of the kernel ${K}$ when $k_{i\perp}^2\to 0$, 
with the corresponding logarithmic UV divergence of the $K$ when $k_{i\perp}^2\to \infty$
removed by a local UVCT($\mu^2)$\cite{Ma:2014jla}.
Use this projection operator, we can sum up all ladder diagrams in Fig.~\ref{fig:quasi-quark} 
in the following symbolic form\cite{Ma:2014jla},
\begin{equation}
\label{eq:fac-co}
\tilde{f}_{q/p}
=
\lim_{m\to\infty} {C}_0 \sum_{i=0}^m {K}^i + \text{UVCT}
=
\left[{C}_0 \frac{1}{1-(1-\widehat{\cal P}){K}}+\text{UVCT} \right]
+ \tilde{f}_{q/p}\, \widehat{\cal P}\, {K}\, ,
\end{equation}
where the term in $[...]$ does not have any CO divergence at the leading power 
of the factorization scale $\mu^2$.  
By combining the terms with $\tilde{f}_{q/p}$, we obtain
\begin{equation}
\tilde{f}_{q/p} = \left[
{C}_0 \frac{1}{1-(1-\widehat{\cal P}){K}} + \text{UVCT} 
\right]\otimes
\left[
\frac{1}{1-\widehat{\cal P}{K}}
\right] ,
\label{eq:fac-q}
\end{equation}
where all leading power CO divergences of the renormalized quasi-quark distribution of a quark of momentum $p$ are now factorized into a ``multiplicative'' factor $[1/(1-\widehat{\cal P}{K})]$, 
which is perturbatively UV finite and equal to the perturbative contribution to the quark distribution.
The same factorization arguments for CO divergences can be applied to the flavor singlet quasi-quark
and quasi-gluon distributions\cite{MaQiu2014}.
\begin{figure}[h]
\begin{center}
\includegraphics[width=4.5in]{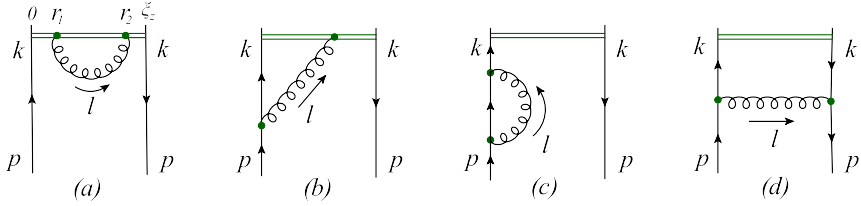}
\caption{\label{fig:oneloop}
Feynman diagrams at one-loop order.}
\end{center}
\end{figure}

\subsection{Renormalization}
\label{sec:renormalization}

To complete the proof of the factorization in Eq.~(\ref{eq:factorization}) 
for the quasi-quark distribution, we need to verify the renormalization of all UV divergences
of the first factor on the right-hand-side of Eq.~(\ref{eq:fac-q}) 
to all orders in $\alpha_s$.

At the first order in $\alpha_s$, the $C_0/(1-(1-\widehat{\cal P}){K})$ in Eq.~(\ref{eq:fac-q})
is given by the Feynman diagrams in Fig.~\ref{fig:oneloop} with all logarithmic CO divergences and corresponding logarithmic UV divergences, specified by the $\widehat{\cal P}K$ factor, removed. 
For quasi-quark distribution, these diagrams still have power UV divergence 
when $l_\perp^2\to\infty$, which requires the additional UVCT, as shown in Eq.~(\ref{eq:fac-q}).
At this order, a UVCT($\mu^2$) to remove the phase space associated with 
the transverse momentum $l_\perp^2 > \mu^2$ is sufficient to renormalize the 
remaining UV divergences of the quasi-quark distribution, 
$\tilde{f}_{q/p}$ in Eq.~(\ref{eq:fac-q})\cite{Ma:2014jla}.  However,
an all order proof of the renormalization of $C_0/(1-(1-\widehat{\cal P}){K})$ 
in Eq.~(\ref{eq:fac-q}) could be more subtle\cite{MaQiu2014}.

\subsection{One-loop matching coefficients}
\label{sec:matching}

Perturbative contribution to the $C_0/(1-(1-\widehat{\cal P}){K})$ in Eq.~(\ref{eq:fac-q})
gives the coefficient function ${\cal C}$ in Eq.~(\ref{eq:factorization}).  
At ${\cal O}(\alpha_s)$, the coefficient function is:
${\cal C}_{q/q}^{(1)}({t},\tilde{\mu}^2,\mu^2,P_z)
=\tilde{f}_{q/q}^{(1)}({t},\tilde{\mu}^2,P_z)-f_{q/q}^{(1)}({t},\mu^2)
$\cite{Ma:2014jla}.
With the $f_{q/q}^{(1)}$ given in the $\overline{\rm MS}$ scheme\cite{Collins:1981uw},
and the $\tilde{f}_{q/q}^{(1)}$ calculated in the transverse momentum cutoff scheme, 
we obtain\cite{Ma:2014jla}, 
\begin{eqnarray}\label{eq:C1}
\frac{{\cal C}_{q/q}^{(1)}({t})}{C_F \frac{\alpha_s}{2\pi}}
{\hskip -0.05in} &=& {\hskip -0.05in}
\left[\frac{1+{t}^2}{1-{t}} \ln\frac{\tilde{\mu}^2}{\mu^2} +1-{t}\right]_+
+ \Bigg[ \frac{t\Lambda_{1-{t}}}{(1-{t})^2}
+ \frac{\Lambda_{{t}}}{1-{t}} +\frac{ \text{Sgn}({t})\Lambda_{{t}}}{\Lambda_{{t}}+|{t}|}
\nnu
&\ & {\hskip -0.2in}
- \frac{1+{t}^2}{1-{t}} \Big[ \text{Sgn}({t})\ln\left(1+\frac{\Lambda_{{t}}}{2|{t}|}\right)
+ \text{Sgn}(1-{t}) \ln\left(1+\frac{\Lambda_{1-{t}}}{2|1-{t}|}\right) \Big]
 \Bigg]_N,
\label{eq:C1qq}
\end{eqnarray}
where $\Lambda_t=\sqrt{\tilde{\mu}^2/P_z^2 + t^2}-|t|$, $\text{Sgn}(t)=1$ if $t\ge 0$, and $-1$ otherwise.  In Eq.~(\ref{eq:C1qq}), the ``+"-function is conventional, and the ``$N$''-function is similarly defined as
\begin{eqnarray}
\int_{-\infty}^{+\infty} d{t} \Big[ g({t})\Big]_N h({t}) =  \int_{-\infty}^{+\infty} d{t}\, g({t}) \left[ h({t}) -h(1)\right],
\end{eqnarray}
where $h({t})$ is any well-behaved function.
As expected, the ${\cal C}_{q/q}^{(1)}$ in Eq.~(\ref{eq:C1qq}), 
so as the one-loop coefficient functions for all other partonic channels, 
are free of any UV, IR and CO divergences\cite{MaQiu2014}.

\section{Summary}

In this talk, we reviewed our proposal for a QCD factorization based approach 
to extract parton distribution and correlation functions from lattice QCD calculations 
of single hadron matrix elements of quark-gluon correlators -- referred as lattice ``cross sections''.  
We presented our strategy to search for good lattice ``cross sections'', 
and their factorization formula to the PDFs.  
As a case study, we discussed in details the quasi-PDFs as lattice ``cross sections''.

\section*{Acknowledgments}

This work was supported in part by the U. S. Department of Energy under Contract No.~DE-AC02-98CH10886 and Grant No.~DE-FG02-93ER-40762,
and the National Science Foundation under Grants No.~PHY-0969739 and No.~PHY-1316617.

\bibliographystyle{ws-ijmpcs}
\bibliography{bibTex1.4}
%

\end{document}